\documentclass[fleqn,twoside]{niaauthm}

\usepackage{graphicx}
\usepackage{amsmath}
\usepackage{xspace}
\usepackage{dcolumn}
\usepackage{bm}
\newcolumntype{d}{D{.}{.}{-1}}

\usepackage{txfonts}  

\newcommand{\HO}{H$_2$O\xspace}
\newcommand{\DO}{D$_2$O\xspace}
\newcommand{\degreesC}{$\,^{\circ}$C\xspace}
\newcommand{\mco}[1]{\multicolumn{1}{c}{#1}}
\newcommand{\SNO}{{\sffamily SNO}\xspace}
\newcommand{\jour}[1]{\textnormal{#1}}

\begin{document}

\begin{frontmatter}

\title{\vspace{-4cm}
       Measurement of \nuc{\bm {222}}{Rn} dissolved in water \\
       at the Sudbury Neutrino Observatory}

\author[Carleton]{I.~Blevis},
\author[Brookhaven]{J.~Boger},
\author[Queens]{E.~Bonvin},
\author[Oxford]{B.~T.~Cleveland},
\author[Oxford]{X.~Dai},
\author[Carleton]{F.~Dalnoki-Veress},
\author[Oxford]{G.~Doucas},
\author[Laurentian]{J.~Farine},
\author[Oxford]{H.~Fergani},
\author[Carleton]{D.~Grant},
\author[Brookhaven]{R.~L.~Hahn},
\author[Queens]{A.~S.~Hamer\thanksref{h}},
\thanks[h]{Deceased.}
\author[Carleton]{C.~K.~Hargrove},
\author[Oxford]{H.~Heron},
\author[Guelph]{P.~Jagam},
\author[Oxford]{N.~A.~Jelley},
\author[Queens]{C.~Jillings},
\author[Oxford]{A.~B.~Knox},
\author[Queens]{H.~W.~Lee},
\author[Carleton]{I.~Levine}$^{,*}$,
\corauth{Corresponding author.  Present address: Department of Physics
         and Astronomy, Indiana University South Bend, South Bend,
         Indiana 46634, USA.}
\ead{ilevine@iusb.edu}
\author[Queens]{M.~Liu},
\author[Oxford]{S.~Majerus},
\author[Queens]{A.~McDonald},
\author[Carleton]{K.~McFarlane},
\author[Carleton]{C.~Mifflin},
\author[Queens]{A.~J.~Noble},
\author[Queens]{S.~No\"{e}l},
\author[Carleton]{V.~M.~Novikov},
\author[Brookhaven]{J.~K.~Rowley},
\author[Carleton]{M.~Shatkay},
\author[Guelph]{J.~J.~Simpson},
\author[Carleton]{D.~Sinclair},
\author[ChalkRiver]{B.~Sur},
\author[Guelph]{J.-X.~Wang},
\author[Brookhaven]{M.~Yeh},
\and
\author[Queens]{X.~Zhu}

\address[Carleton]{Carleton University, Ottawa, Ontario K1S 5B6,
                   Canada}
\address[Brookhaven]{Chemistry Department, Brookhaven National
                     Laboratory, Upton, New York 11973-5000, USA}
\address[Queens]{Department of Physics, Queen's University, Kingston,
                 Ontario K7L 3N6, Canada}
\address[Oxford]{Department of Physics, University of Oxford, Denys
                 Wilkinson Building, Keble Road, Oxford, OX1 3RH, UK}
\address[Laurentian]{Department of Physics and Astronomy, Laurentian
                     University, Sudbury, Ontario P3E 2C6, Canada}
\address[Guelph]{Physics Department, University of Guelph, Guelph,
                 Ontario  N1G 2W1, Canada}
\address[ChalkRiver]{Atomic Energy of Canada Limited, Chalk River
                     Laboratories, Chalk River, Ontario K0J 1J0, Canada}

\begin{abstract}

The technique used at the Sudbury Neutrino Observatory (\SNO) to
measure the concentration of \nuc{222}{Rn} in water is described.
Water from the \SNO detector is passed through a vacuum degasser (in
the light water system) or a membrane contact degasser (in the heavy
water system) where dissolved gases, including radon, are liberated.
The degasser is connected to a vacuum system which collects the radon
on a cold trap and removes most other gases, such as water vapor and
N$_2$.  After roughly 0.5~tonnes of \HO or 6~tonnes of \DO have been
sampled, the accumulated radon is transferred to a Lucas cell.  The
cell is mounted on a photomultiplier tube which detects the
$\alpha$-particles from the decay of \nuc{222}{Rn} and its progeny.
The overall degassing and concentration efficiency is about 38\% and
the single-$\alpha$ counting efficiency is approximately 75\%.  The
sensitivity of the radon assay system for \DO is equivalent to
$\sim$3$\times 10^{-15}$ g~U/g~water.  The radon concentration in both
the \HO and \DO is sufficiently low that the rate of background events
from U-chain elements is a small fraction of the interaction rate of
solar neutrinos by the neutral current reaction.

\end{abstract}


\end{frontmatter}

\section{Introduction}
\label{sec:intro}

The Sudbury Neutrino Observatory (\SNO) is a heavy water Cherenkov
detector which was built to understand why all previous solar neutrino
experiments
\cite{bib:solar_experiments,bib:kamiokande,bib:sage,bib:gallex,bib:sk,%
bib:gno} have observed fewer neutrinos than are predicted by generally
accepted solar models \cite{bib:sm1,bib:sm2}.

The \SNO detector is described in detail elsewhere \cite{bib:SNONIM}.
Briefly, \SNO consists of an inner neutrino target of 1000~tonnes of
ultra-pure D$_2$O contained in a 12~m diameter spherical, transparent,
acrylic vessel.  An array of 9438 photomultiplier tubes, mounted on an
$\sim$18~m diameter stainless steel geodesic support structure, detect
the Cherenkov light from electrons produced by neutrino interactions
in the \DO.  The volume between the acrylic vessel and the tube
support structure contains approximately 1700~tonnes of ultra-pure
H$_2$O which shields the \DO volume from high-energy $\gamma$~rays
produced by radioactivity in the outer regions of the detector.

Outside the photomultipliers lie an additional 5700~tonnes of \HO
shielding.  The two water shielding regions are separated by a nearly
impermeable water seal which serves to keep the water in the outer
shielding area, which has higher radon levels, isolated from the water
between the photomultipliers and the acrylic vessel, where the radon
level is lower.  The external water shielding region is viewed by
91~outward-looking photomultiplier tubes which help to reject
background from muons traversing the detector.  The detector is
situated in the INCO, Ltd.\ Creighton mine, in Sudbury, Ontario,
Canada.  At a depth of 6800 feet, only about 70~muons interact in the
detector per day.

\markboth {\hfill Radon assay method \hfill}
          {\hfill Radon assay method \hfill}

\SNO detects solar neutrinos through three distinguishable
interactions with the D$_2$O target:
\begin{flalign*}
\nu_\text{e}\mspace{2.5mu} & + \text{d} \       \longrightarrow
        \text{e}^- \!+ \text{p} + \text{p} & \text{(CC),} \\
\nu_x                      & + \text{d} \       \longrightarrow
        \nu_x        + \text{p} + \text{n} &  \text{(NC),} \\
\nu_x                      & + \text{e}^-  \!\! \longrightarrow
        \nu_x        + \text{e}^-          &  \text{(ES),}
\end{flalign*}
where $x$ denotes any of the active neutrino species $\text{e}, \mu,
\text{or } \tau$.  The CC reaction is only sensitive to the flux of
electron neutrinos, whereas the NC reaction is equally sensitive to
all active neutrino flavors.  Three techniques have been developed to
observe the neutrons from the NC reaction in \SNO.  In the first phase
of the experiment, the inner vessel was filled with pure \DO.  The
neutrons were captured by deuterium nuclei creating 6.25-MeV
$\gamma$~rays which interacted to make relativistic electrons whose
Cherenkov light was detected.  In the second phase, NaCl was added to
the D$_2$O.  Most neutrons then capture on $^{35}$Cl, an exothermic
reaction that yields photons whose energies sum to 8.6~MeV.  In the
third phase, the salt will be removed and neutrons will be detected
with low-background $^3$He-filled counters that will be installed in
the \DO.

The first publication \cite{bib:SNOP1} of \SNO results indicated that
the flux of electron neutrinos with energy $\ge$6.75~MeV inferred from
the CC reaction is not as large as the total rate inferred from the ES
reaction as measured by \SNO, or with greater accuracy, by
Super-Kamiokande \cite{bib:sk}.  Since the ES reaction is mostly
sensitive to the flux of electron neutrinos, but has a small
contribution from the flux of other neutrino flavors, this implies
that the flux of active neutrinos from the Sun is greater than the
observed flux of electron neutrinos alone.  As only electron neutrinos
are produced in the Sun, this observation is evidence that electron
neutrinos have transformed into some combination of $\mu$ and $\tau$
neutrinos by the time they reach the detector.

A much higher precision measurement of this phenomenon was obtained by
a comparison at energies $\ge$5~MeV of the CC and NC rates in the \SNO
detector \cite{bib:ccnc} and their temporal variations \cite{bib:dn}.
To make this comparison required that the radioactive backgrounds in
the detector were well understood, since decays of progeny of
$^{238}$U (``U-chain'') and progeny of $^{232}$Th (``Th-chain'') can
mimic the neutrino interactions.  The isotopes of most concern for the
CC/NC comparison are \nuc{214}{Bi} in the U-chain and \nuc{208}{Tl} in
the Th-chain because their decays can produce $\gamma$~rays with
energies greater than 2.2~MeV.  These high-energy $\gamma$~rays can
photodisintegrate the deuteron, producing a free neutron, and thus
mimic the NC disintegration of the deuteron.

Two techniques, which are discussed elsewhere \cite{bib:suf,bib:mnox},
have been developed to measure the aqueous concentration of $^{226}$Ra
from the U-chain and $^{224}$Ra from the Th-chain.  These give a good
measurement of the concentration of radium ions in the water.
Knowledge of the $^{226}$Ra concentration is, however, not sufficient
to determine the U-chain radioactive background because $^{222}$Rn,
the decay product of $^{226}$Ra, is a noble gas with a 3.8-d
half-life.  Small leaks of air and traces of radium in detector
materials can introduce $^{222}$Rn, and lead to significant
disequilibria between $^{226}$Ra and $^{214}$Bi.  To properly
understand the radioactive background, it is thus imperative to
measure directly the $^{222}$Rn concentration in the water as only
isotopes with short half-lives separate \nuc{222}{Rn} from the
undesired $^{214}$Bi.

The underground air at the \SNO laboratory contains $\sim$100~Bq/m$^3$
of \nuc{222}{Rn} (3~pCi/l or 50~Rn atoms/cm$^3$), which, if the radon
were at its equilibrium concentration with the \DO, would yield a
dissolved \nuc{222}{Rn} level almost $10^6$ times higher than
tolerable.  Considerable precautions were thus taken in the design and
construction of \SNO to limit the leakage of \nuc{222}{Rn} into the
detector.  As examples of such measures, all components of the water
systems were selected for low radon diffusion and emanation, the
entire \DO system was leak checked with a He mass spectrometer, and
one of the final elements of both the \HO and \DO water purification
systems is a degasser that has a radon removal efficiency of $>$90\%.
Further, the polypropylene pipes in the water system have thick walls
and are made with specially selected low-radioactivity material
\cite{bib:himont}, all detachable joints are sealed with butyl rubber
O-rings \cite{bib:butyl} which have low radon permeability and
emanation, the detector cavity is lined with a membrane with low Rn
permeability, and there is a ``cover gas'', a continuously flowing
stream of N$_2$ from the boiloff of liquid nitrogen, into the vapor
space directly above the \DO and \HO.


\section{Maximum allowable radon concentration}

The standard model calculation of the $^8$B neutrino flux from the
Sun~\cite{bib:sm1}, together with the cross section for the NC
reaction \cite{bib:ncxsec}, predicts that 10~to 15~neutrons will be
produced by solar neutrinos per day in the 1000~tonnes of \DO in the
\SNO detector.  Based on this prediction, \SNO set the design goal to
have no more than one neutron per day produced in the \DO by U-chain
contamination.  As described in \cite{bib:mnox}, this implies that the
maximum allowable contamination of the D$_2$O is $3 \times
10^{-14}$~g~U/g~D$_2$O, assuming secular equilibrium between $^{238}$U
and $^{214}$Bi.  This can be translated into a maximum concentration
for $^{222}$Rn in the D$_2$O of about 0.2~atoms/liter or
0.4~mBq/m$^3$.  The maximum allowable radon concentrations in the \HO
between the acrylic vessel and the photomultipliers and in the outer
bulk shielding water are given in Table~\ref{tab:goals}.  At these
levels, the background contributed by \HO to the NC signal is no more
than that from the \DO.

\begin{table}[htb]
\caption{Maximum allowable radon concentration in the three detector
water regions, the D$_2$O, the inner H$_2$O region between the acrylic
vessel and the photomultipliers (I), and the outer H$_2$O region
between the photomultipliers and the cavity wall (O).  The fourth
column is the U concentration assuming secular equilibrium with
$^{222}$Rn.}
\label{tab:goals}
\begin{tabular*}{\hsize}{@{} l @{\extracolsep{\fill}} d d @{\extracolsep{-2em}} d @{}}
\hline
\hline
Region     & \mco{Atoms Rn/liter} & \mco{mBq Rn/m$^3$} & \mco{\hspace{4em}g U/g water}  \\
\hline
D$_2$O     & 0.2                  &  0.4               & 3.0 \times 10^{-14} \\
H$_2$O (I) & 3                    &  6                 & 4.5 \times 10^{-13} \\
H$_2$O (O) & 7                    & 14                 & 1.1 \times 10^{-12} \\
\hline
\hline
\end{tabular*}
\end{table}

To reduce the systematic error in the NC determination that arises
from the uncertainty in the radon level to less than a few percent, it
is necessary to lower the radon content of the water below the maximum
allowable level.  Such levels can only be measured with accuracy if
the ultimate sensitivity of the assay method is an order of magnitude
below the design goal.  As shown below, these requirements were met,
thus allowing the critical neutral current measurement to be made with
precision.

\section{Assay of the H$_{\bm 2}$O}
\label{sec:h2o}

\subsection{Overview of the assay technique}

The assay method operates by pumping water from the detector to a
degasser which extracts the radon.  The radon is collected,
concentrated, and then transferred to a miniature ZnS-coated
scintillation counter (a Lucas cell~\cite{bib:lucas}) for measurement.
The extraction of radon is performed differently in the H$_2$O and
D$_2$O systems; we describe the extraction from H$_2$O here and from
D$_2$O in Sec.~\ref{sec:d2o}.  The radon collection and counting are
the same in both systems and are discussed in Sec.~\ref{sec:rncollect}
and Sec.~\ref{sec:counting}, respectively.

\subsection{Water flow}

\begin{figure}
\includegraphics[width=\hsize]{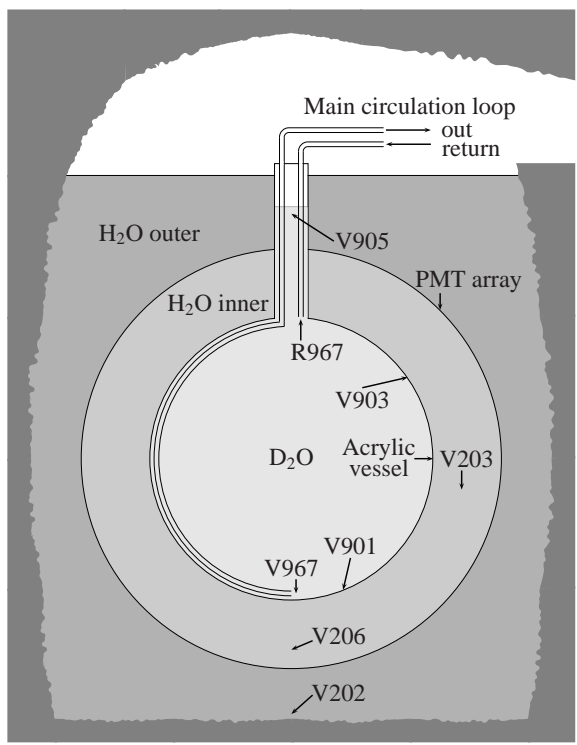}
\caption{Location of sample points for usual radon assays of H$_2$O
and D$_2$O.  Assay points are labelled by the valves that must be
opened to extract the water.  Most D$_2$O assays take water from the
main purification loop shown on the left side of the acrylic vessel.
This circulation normally draws water from the bottom of the vessel,
and, after purification and assay, returns it at the bottom of the
vessel neck.  Flow in the reverse direction is also possible.
Measurements of radon levels are also made occasionally at other
locations within the vessel (V901, V903, and V905).  Assays of the
light water are usually conducted at two positions between the
photomultiplier array and the acrylic vessel (V203 and V206) and at
the bottom of the cavity (V202).}
\label{fig:sample}
\end{figure}

H$_2$O can be drawn from six locations in the detector.  Two sample
points are at the bottom of the cavity, three at the equator of the
acrylic vessel, and one at the bottom of the photomultiplier array.
The most commonly sampled H$_2$O and D$_2$O positions are shown in
Fig.~\ref{fig:sample}.  The distance from the sample points to the
degasser is about 70~m and the piping is polypropylene with 60~mm
outer diameter and 5.5~mm wall thickness.  Polypropylene was chosen
for its low radioactivity, low leaching of impurities in the presence
of ultra-pure water, and low radon permeability.  The water sample is
taken by a diaphragm pump \cite{bib:diaphragm} whose wetted portions
are made from polypropylene, except for the diaphragm which is Teflon.
A typical assay samples water for 30~minutes at a flow rate of
19~liters/minute.  At the maximum allowable contamination of the
H$_2$O between the acrylic vessel and the photomultipliers ($4.5
\times 10^{-13}$~g~U/g~\HO), this water would contain 1480~$^{222}$Rn
atoms.  The water goes through a vacuum-degassing chamber and is
returned to the main H$_2$O circulation system by another diaphragm
pump.


The volume of water is measured with either a rotameter flowmeter or a
stroke counter attached to the diaphragm pump.  The volume per stroke
is 0.34~liters, and is independent of flow rate over the range of
flows normally used.  Both of these were calibrated by flowing water
from the pump to a container on a scale.  One of the larger sources of
systematic uncertainty is the difficulty of accurately measuring the
water flow rate from the pulsating diaphragm pump.  The estimated
uncertainty is 14\% in the flowmeter readings and 10\% in the
stroke-counted readings.

\subsection{The light water monitor degasser}

\begin{figure}
\begin{center}
\includegraphics[width=0.8\hsize]{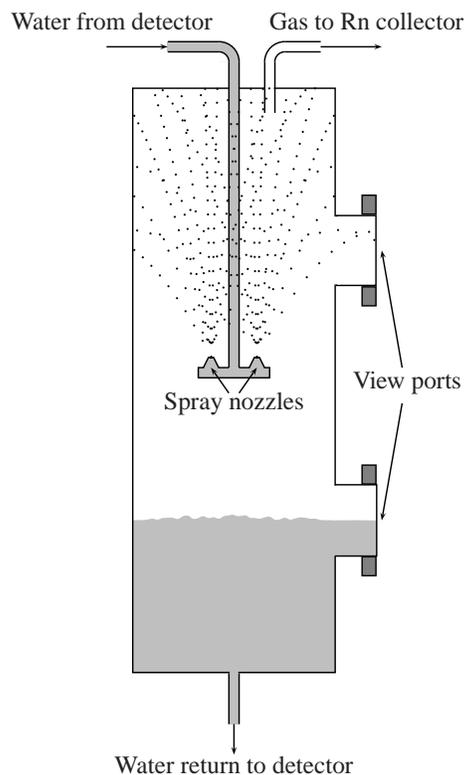}
\caption{Schematic diagram of the monitor degasser in the H$_2$O system.}
\label{fig:degasser}
\end{center}
\end{figure}

The monitor degasser (MDG)
\cite{bib:liu_thesis,bib:zhu_thesis,bib:noel_thesis} is shown in
Fig.~\ref{fig:degasser}.  It is a custom-designed stainless steel
vertical cylinder 1.15~m high and 0.4~m in diameter
\cite{bib:allweld}.  The usual pressure difference between the feed
water line and the degassing chamber is 45~PSI.  The water enters at
the top of the degasser at a temperature of 13\degreesC and passes
through three stainless-steel full-cone spray nozzles
\cite{bib:nozzles}.  The nozzles spray upwards and produce water
droplets with a diameter of $\sim$1~mm.  As the drops fall, or run
down the walls of the degasser chamber, the dissolved gases leave the
water and are drawn off to the radon collector described in
Sec.~\ref{sec:rncollect}.  The spray and the water level are visible
through two acrylic view ports mounted on flanges welded to the side
of the vessel.  The water level in the bottom of the degasser is
maintained at a height of $\sim$0.3~m.

The degassing efficiency of the MDG was measured by bringing a small
volume of water into equilibrium with radon in the air of the
underground laboratory, injecting this radon-enriched water into the
input of the MDG, and then extracting, concentrating, and counting the
radon.  The number of detected radon atoms was compared to the number
expected based upon the radon solubility, the known radon activity in
the air (measured by introducing a different sample of the same air
into an evacuated Lucas cell and counting its activity), and the known
efficiencies of collection and concentration.  With H$_2$O at
13\degreesC and a flow rate of 19~liters/minute the degassing
efficiency was measured to be $0.58\pm0.10$~\cite{bib:wrightson}.

\subsection{Assay system background}
\label{sec:h2oback}

Although the materials in the assay system were selected for their low
diffusion and emanation rates for radon \cite{bib:emanation}, some
radon can still enter the water being assayed through leaks, by
diffusion through pipes, O-rings, etc., by emanation from contaminants
in assay system components, and by emanation from surface dust,
embedded dirt, etc.

The background of the degasser was measured by filling the MDG with
water and flowing this water in ``closed loop mode'' (which sends the
output water of the degasser back to its input) until it was
completely degassed.  Then several assays of the degassed water were
conducted, each of 3-hour duration, at a flow rate of
20~liters/minute.  In these experiments $19 \pm 4$~atoms of radon
entered the system per hour of assay by the combination of leakage,
diffusion, and emanation.  For a typical 30-min assay this represents
a background of 1.1\% of the number of atoms that would be extracted
and collected if the water were at the maximum allowable level.  These
measurements, however, were only of the MDG and the subsequent radon
collection apparatus and did not include the assay system piping that
leads to the detector.  Furthermore, the background can change as a
function of time.  For instance, we find that vibrations may loosen
the nuts on valves and flanges, for which we must compensate by
periodic tightening.  Thus, the background for any given assay may be
higher than measured in these closed loop assays, where great care was
taken to ensure that the system was tightly sealed.  To take account
of these additional sources of background, some of which may vary in
time, we add onto the previously stated statistical error of
$\pm4$~atoms/hr an additional systematic error, which we estimate to
be $^{+8}_{-4}$~atoms/hr.  The total assay system background is thus
$19^{+9}_{-6}$~atoms/hr.


\section{Assay of the D$_{\bm 2}$O}
\label{sec:d2o}

The basic principles of the assay system for D$_2$O are very similar
to those for H$_2$O, namely, degassing followed by radon collection,
concentration, and counting.  The D$_2$O system, however, must be an
order of magnitude more sensitive than the H$_2$O system (see
Table~\ref{tab:goals}), be able to function in 0.2\%~NaCl-D$_2$O
brine, and, since the D$_2$O is so valuable, have minimal loss of
D$_2$O vapor.  For these reasons a polypropylene hydrophobic-membrane
contact degasser was chosen, rather than a metal vacuum degasser.  The
system design and calibration is described in more detail in
\cite{bib:darren}.

\subsection{Water flow}

The D$_2$O can be assayed from seven locations in the detector as well
as at points within the water purification system itself.  The most
commonly sampled points are labeled in Fig.~\ref{fig:sample}.  There
are five assay positions within the D$_2$O on the acrylic vessel wall:
at the bottom of the vessel (V901), 1/3 of the way up (not shown), 3/4
of the way up (V903), at the bottom of the neck (not shown), and 0.3~m
below the surface of the water in the neck (V905).  In addition, the
water from the main purification circulation loop can be sampled
either at the bottom of the neck (R967) or at the bottom of the vessel
(V967).  The pipes within the vessel are made from the same acrylic as
the vessel itself; outside the vessel the pipes are polypropylene, as
in the H$_2$O system.  For the dedicated assay lines, a diaphragm pump
identical to that used in the H$_2$O system draws water from the
vessel to the D$_2$O degasser.

A typical assay samples water for 5~hours at 21~liters/min.  At the
maximum allowable level this water contains about 1260~\nuc{222}{Rn}
atoms, which, taking into account the degassing efficiency and
transfer efficiency, results in 480~atoms in the Lucas cell at the end
of an assay.  Since the single-$\alpha$ counting efficiency is 74\%
and 3~prompt alphas are emitted per radon decay (see
Sec.~\ref{sec:LCeff}), this gives about 180~events in the first day of
counting.  This should be compared with the typical Lucas cell
background rate of about 10~counts/d, and the background from the
assay system, which contributes about 20~counts in the first day of
counting.  Defining the sensitivity as when the signal equals
approximately three times the uncertainty of the background
\cite{Currie}, the sensitivity of the entire D$_2$O assay system in
the current configuration is about one-tenth of the maximum allowable
level, or about $3 \times 10^{-15}$~g~U/g~D$_2$O.


There is no flow meter in the D$_2$O radon assay system.  Rather, the
flow rate is set by adjusting the pump stroke rate to a fixed value.
Since the D$_2$O system uses the same type of diaphragm pump as the
H$_2$O system, we use its calibration, whose uncertainty is 10\%.
There may, however, be small differences in any two pumps believed to
be identical, and thus we add in quadrature an assumed 5\%
uncertainty, giving a total flow rate uncertainty in stroke-counted
experiments of 11\%.  Some assays were conducted before the
installation of the stroke counter; for these we estimate a 17\%
uncertainty in the flow rate.

\subsection{The heavy water monitor degasser}

The MDG in the D$_2$O system is a membrane contact degasser
\cite{bib:celgard}.  It consists of bundles of hollow, porous,
hydrophobic polypropylene fibers woven around a hollow polypropylene
water distribution tube, all of which is contained in a polypropylene
housing.  The distribution tube is plugged at the center, and a baffle
in the containment cartridge forces the water to flow to the outside
of the cartridge and pass over the tightly packed fibers.  As the
water flows over the fibers, the dissolved gases pass through the
fiber walls to their hollow center from which a vacuum pump draws the
gases into a radon collection system similar to that in the H$_2$O
system.  The water, on the other side of the baffle, goes back to the
central water tube, exits the degasser, and is returned to the main
D$_2$O purification system.

The degassing efficiency of the \DO MDG for radon has not been
measured directly, but we can infer its efficiency from other
experiments.  By measuring the radon concentration of the water that
enters and exits the \DO process degasser, its efficiency was found to
be $83 \pm 5\%$ at a flow rate of 195~liters/min.  The process
degasser contains two parallel sets of three membrane degassers in
series where each degasser is of identical construction to the one in
the MDG, so the inferred efficiency of a single degasser cell for
radon is $45 \pm 3\%$ at 97.5~liters/min.  This flow rate is much
higher than the rate through the MDG which is customarily
21~liters/min.  To extrapolate to lower flow rate, we can use an
approximate membrane degasser model~\cite{bib:sengupta} which predicts
that the efficiency $\varepsilon_{\text{degas}}$ varies with flow rate
$F$ as $\ln(1 - \varepsilon_{\text{degas}}) \propto F^{-\chi}$ where
$\chi$ depends on degasser module geometry but is independent of gas
species.  The value of $\chi$ can be found for our degasser by
applying this equation to measurements of the oxygen degassing
efficiency of the process degasser at different flow rates.  A
probe~\cite{bib:honeywell} with ppb sensitivity in high resistivity
liquids such as ultra-pure D$_2$O was used, and the measured oxygen
degassing efficiency of one cell of the process degasser was $84 \pm
3\%$ at 21~liters/min and $67.9 \pm 2.4\%$ at 97.5~liters/min.  From
these measurements we infer $\chi = 0.31$, and, using the measured
radon efficiency at 97.5~liters/min, the predicted degassing
efficiency of the MDG for radon is 62\% at 21~liters/min.  Since
several assumptions are involved in this model which may not be
completely satisfied, such as independence of the degassing efficiency
on the gas concentration, we assign a liberal systematic uncertainty
to this estimate.  The upper limit is set by noting that the
efficiency for radon can be no more than that for oxygen at the same
flow rate, i.e., it must be less than 84\%.  A firm lower bound for
the radon degassing efficiency at 21~liters/min is at 45\% as that was
the measured efficiency at 97.5~liters/min.  We consider these extreme
bounds to be effective two sigma uncertainties.  Our estimate for the
radon degassing efficiency of the \DO MDG is thus
$62^{+11}_{-\phantom{1}9}\%$.


\subsection{Assay system background}
\label{sec:d2oback}

The emanation and leak background of the MDG and radon collector were
measured under static conditions \cite{bib:darren} as follows: The MDG
and collector were sealed and their helium leak rate was measured to
be less than $10^{-8}$~cm$^3$/s.  Then the system was evacuated and
isolated for as long as 11~days, at the end of which the gas in each
evacuated component was individually collected and counted.  The most
significant background was $16 \pm 1.5$~Rn atoms/hr and came from the
trap used to remove water vapor from the gas stream (see
Fig.~\ref{fig:rnboard}).  The only other appreciable background was
from the degasser portion of the system which contributed $1.7 \pm
0.2$~\nuc{222}{Rn} atoms/hr.  Adding these two components gives a
total assay system background of $18 \pm 1.5$~\nuc{222}{Rn} atoms/hr,
which, for a typical assay time of 5~hours, gives a background of $90
\pm 8$~\nuc{222}{Rn} atoms on the collection trap at the end of the
assay.  This is much less than the $\sim$780~atoms that would be
collected on the trap from the water if it were at the maximum
allowable level.


The concerns regarding possible changes in background with time that
were expressed for the H$_2$O system also pertain here.  To account
for such time-dependent changes in the background, we add, as for the
\HO system, a systematic uncertainty of $^{+8}_{-4}$~atoms/hr, which
makes the total assay system background $18^{+8}_{-4}$~atoms/hr.

\section{The radon collector}
\label{sec:rncollect}

\begin{figure*}
\includegraphics[width=\hsize]{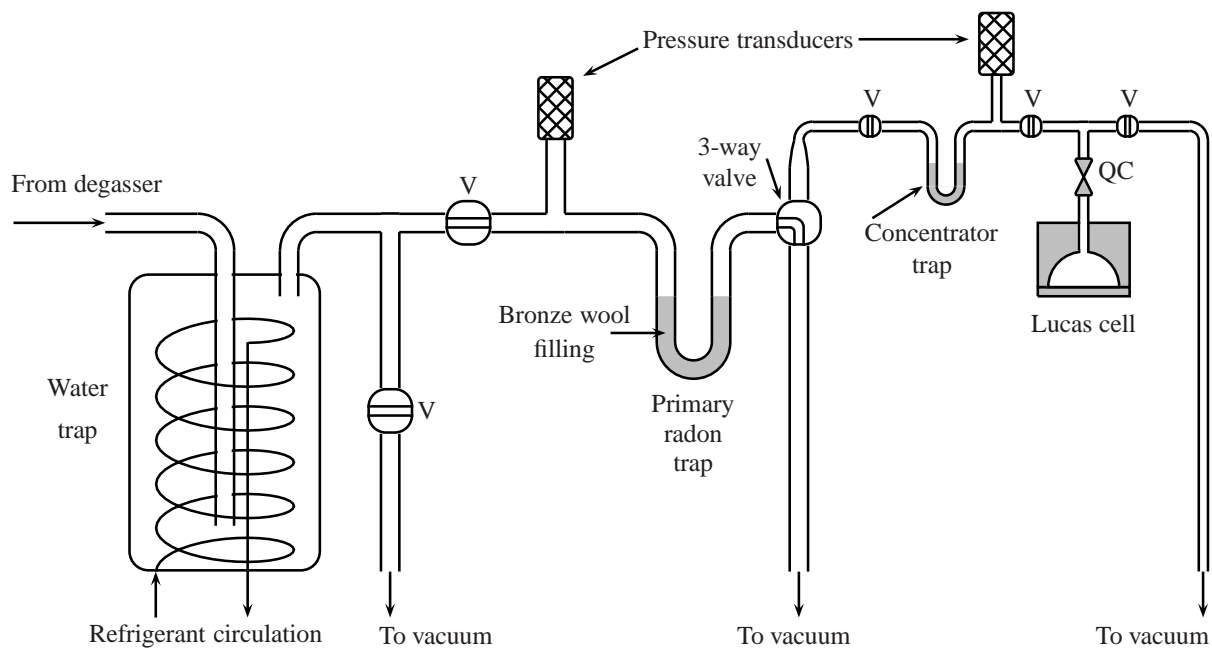}
\caption{Schematic of radon collection and concentration apparatus.
Parts are not drawn to scale.  ``V'' shows a valve and ``QC'' labels
the quick-connector on the Lucas cell.  The valves are drawn in the
position for extraction from the degasser, during which time the
primary radon trap is immersed in liquid N$_2$.}
\label{fig:rnboard}
\end{figure*}

We describe here how the radon that is extracted from the water by the
monitor degasser is collected, separated from other gases, and
concentrated.  The equipment used for this purpose is nearly the same
in both the H$_2$O and D$_2$O systems and is illustrated in
Fig.~\ref{fig:rnboard}.  It is based on apparatus developed to measure
radon emanation for materials selection during the \SNO design phase
\cite{bib:emanation}.

In overview, the gases extracted from the water are first drawn by a
vacuum pump through a cold trap which removes water vapor.  The gas
stream then flows through a liquid nitrogen cooled trap filled with
bronze wool which stops radon, but allows N$_2$ and O$_2$ to pass
through.  At the end of the sampling period the captured radon is
transferred to a concentrator trap and then to a specially-developed
low-background Lucas cell.  These various components and their use in
an assay will now be described.

\subsection{Collector components and use}

The gas stream first enters a water vapor trap~\cite{bib:fts}.  In the
H$_2$O system, where the vapor load is about 10~cm$^3$/min, the trap
is an acrylic cylinder with a volume of 42~liters inside of which is a
refrigeration coil.  The coil is held at -60\degreesC during an
extraction.  The D$_2$O system has a similar device but it is smaller
as the vapor load is less by about a factor of 10.

After passing through the water trap, the dry gas enters the primary
radon trap which consists of a 10~mm diameter stainless steel tube
stuffed with cleaned bronze wool and bent into a `U' shape.  This
tube, whose volume is $\sim$50 cm$^3$, is immersed in liquid nitrogen
during the extraction.  N$_2$, O$_2$, and Ar in the gas stream go
through this trap, but radon, CO$_2$ and any residual water vapor
adhere to the cold bronze wool.  When a sufficient sampling time has
elapsed, the valve at the inlet to the radon collector is closed.
Evacuation of the trap is continued for a few minutes to remove traces
of nitrogen and oxygen.  Next, the valve to the vacuum pump is closed
and the valve between the primary trap and the (previously evacuated)
concentrator trap is opened.  The concentrator trap is a coiled tube
of 3~mm stainless steel tubing with 300~mm length that is also stuffed
with bronze wool.  The primary trap is heated with a heat gun and the
concentrator trap is chilled with liquid nitrogen so the gas is
cryopumped to the concentrator trap.

The cryopumping continues for about 15~min during which time a Lucas
cell is attached to the system at an adjacent quick-connect port and
evacuated.  The concentrator trap is then isolated and heated and a
valve between the Lucas cell and the concentrator trap is opened.
Since the volume of the Lucas cell is larger than that of the
concentrator trap and connecting tubing, most of the radon transfers
to the Lucas cell by volume sharing.  The pressure in the cell at the
time of filling is read with a transducer and is typically $<$0.1~atm
for \HO assays or $<$0.5~atm for \DO assays.  This gas is mainly
CO$_2$ which is not well separated from radon as they have similar
boiling temperatures.  Even at 0.5~atm, however, the $\alpha$-particle
range is nearly 6 cm, considerably greater than the 1.9 cm radius of
the Lucas cell.  Finally, the cell is disconnected and taken to a
counting facility on the surface.


Except for the water trap, all components of the radon collector are
off-the-shelf stainless steel parts connected with compression
fittings.  Considerable care was taken in the section from the
concentrator to the Lucas cell to minimize the volume by using small
diameter tubing, filling all unused volumes with inert material, and
selecting low dead-volume valves.

The efficiency of the primary trap for stopping radon was measured
under static conditions by concentrating a large amount of radon from
the air into a Lucas cell, counting the radon in the cell, injecting
the radon from that cell into the input of the primary trap, and
extracting the radon back into the same cell.  After accounting for
radon decay, radon absorption into the acrylic of the Lucas cell, and
the transfer efficiency from the concentrator trap to the Lucas cell,
the trapping efficiency was found to be $100.5 \pm 2.3\%$.

The efficiency of transfer from the concentrator trap to the Lucas
cell was measured in the H$_2$O system by filling the concentrator
trap with air at atmospheric pressure.  An evacuated Lucas cell was
then attached and the change in pressure gave an efficiency of $64 \pm
2\%$.  For the D$_2$O system this technique yielded an efficiency of
$63.8 \pm 2.0\%$.

An alternate efficiency measurement was made for the D$_2$O system
using radon.  A Lucas cell containing a measured amount of radon was
attached to the usual Lucas cell port and connected to the
concentrator trap.  The concentrator trap was chilled with liquid
nitrogen and the radon from the cell was drawn by a vacuum pump to the
concentrator trap.  After pumping all radon from the cell, the trap
was heated and the radon was re-injected into the same Lucas cell by
the usual volume-sharing technique.  After accounting for radon decay,
the transfer efficiency from the concentrator trap to the Lucas cell
was measured to be $61.8 \pm 1.0\%$.  This more precise value is used
to infer the radon concentration in the D$_2$O.

\subsection{Radon collector background}

The background of the radon collector water trap is included in the
total assay system background and was considered above in
Sec.~\ref{sec:h2oback} and Sec.~\ref{sec:d2oback}.  The background of
the other components of the radon collector in the D$_2$O system was
measured \cite{bib:darren} by helium leak testing to
$10^{-8}$~cm$^3$/sec, evacuation of the entire collector, and then
closing all valves so as to isolate the separate parts.  After a seal
time of a few days, the individual regions were each extracted
independently into Lucas cells.  Weighting each section of the
collector by the time it is used in the processing of the assay gives
a total background that is several times less than the assay system
background.  We therefore consider the background of the collector
(exclusive of the water trap) to be negligible.

\section{Data acquisition system}
\label{sec:counting}

The Lucas cell from the assay is placed on the end of a 50-mm diameter
10- or 12-stage photomultiplier tube and counted for 8~to 10~days.
The tube output is amplified, digitized, and stored in 256~channels of
a 4096-channel analyzer.  Every 3~hours the data stored in the
analyzer are transferred to a computer and a ``log'' file is updated
with the cumulative counting time and the cumulative number of events
within a chosen region of interest.  Another file is also
written every three hours which contains the full cumulative energy
spectrum.  If desired, these spectra can be reanalyzed with the region
of interest
redefined and a new ``log'' file generated.  The \nuc{222}{Rn} decay
counting system contains 11~counting stations, with one or two Lucas
cells assigned to each station.  Since the system is in a laboratory
on the surface, there is a delay of two or more hours between the end
of an assay and the start of counting.

\subsection{Lucas cells}

The Lucas cells developed for \SNO are acrylic cylinders with a hollow
interior machined into a 19-mm radius hemisphere whose surface is
painted with activated ZnS.  One end of the cylinder has a low-volume
quick-connector \cite{bib:swagelok} through which radon gas is
admitted and the other end is sealed with a flat sheet of acrylic.
When $\alpha$ particles from the decay of $^{222}$Rn or one of its
$\alpha$-emitting decay products strike the ZnS coating on the acrylic
surface, light is emitted.  The flat acrylic end of the cell is placed
atop a photomultipler tube which detects this scintillation light.
The cell diameter is 5~cm and the interior gas volume is 15.5~cm$^3$.
These low background devices are described in detail in
\cite{bib:emanation}.

New cells are leak checked with a He mass spectrometer and the cells
in use are periodically checked by evacuating them underground,
keeping them for several hours in this high radon environment, and
bringing them to the surface for counting.  A cell filled with
underground air contains about 900~Rn atoms, compared to $\sim$480
from D$_2$O at the maximum allowable level.

One cause of leakage is improper removal of the Lucas cell from the
assay system at its quick-connector at the end of an extraction.  This
problem appears to have occurred once in the assays of the D$_2$O.
Leakage can also occur because of inadequate lubrication of, or dirt
on, the connector O-rings.  This problem was greatly reduced by
instituting a regular program of disassembly, cleaning, and
regreasing.  This is essential because even small leaks can introduce
a number of radon atoms comparable with the $\sim$150~atoms that are
presently collected in a typical \DO assay.


\subsubsection{Lucas cell efficiency}
\label{sec:LCeff}

The section of the U chain that begins with \nuc{222}{Rn} is
  \nuc{222}{Rn} $\xrightarrow[\text{3.82 d}]{\text{5.5 MeV }\alpha}$
  \nuc{218}{Po} $\xrightarrow[\text{3.10 m}]{\text{6.0 MeV }\alpha}$
  \nuc{214}{Pb} $\xrightarrow[\text{26.8 m}]{\beta}$
  \nuc{214}{Bi} $\xrightarrow[\text{19.9 m}]{\beta}$
  \nuc{214}{Po} $\xrightarrow[\text{162 $\mu$s}]{\text{7.7 MeV }\alpha}$
  \nuc{210}{Pb}, which has a 22-year half life.
Shortly after the decay of a \nuc{222}{Rn} atom two additional
$\alpha$ particles are emitted, from \nuc{218}{Po} and \nuc{214}{Po}.
The effective efficiency of the Lucas cell for \nuc{222}{Rn} decay is
thus three times its efficiency for single-$\alpha$ detection,
provided counting begins a few hours or more after the cell is filled.

The efficiency of the \SNO Lucas cells was measured by injecting air
containing a known amount of radon into two cells at a commercial
radon calibration company \cite{bib:pylon}.  The company measured the
radon concentration of the air put into the cells to be $1649 \pm 66
\text{ (stat)} \pm 12 \text{ (syst)}$~Bq/m$^3$, equivalent to about
12~000~Rn atoms in each cell.  The two cells were counted at \SNO
about one day after the radon injection.  After accounting for decay
between the time of injection and the start of counting, the cells
were found to have single-$\alpha$ detection efficiencies of 75\% and
74\% with a 0.6\% statistical uncertainty and a 3.4\% systematic
uncertainty.


Soon after the calibration of these two cells, nine Lucas cells were
taken underground, where the typical radon concentration is
100~Bq/m$^3$, evacuated, and filled with ambient air.  All cells were
found to have similar efficiency but with a statistics dominated 12\%
uncertainty in each measurement.  Assuming all cells have the same
efficiency, the standard deviation of efficiency from the combination
of these measurements is $\pm$7\%.  Combining this with the known
difference in efficiency from cell to cell due to their volume
difference, which is $<$3\%, we conclude that the single-$\alpha$
detection efficiency of cells that were not directly calibrated is $74
\pm 7\%$.  The 26\% loss of $\alpha$-particles is mainly due to the
cell geometrical efficiency.


\subsubsection{Lucas cell background}

When a Lucas cell is new, its background from cosmic radiation and
radioactivity in the ZnS is less than 3~counts per day, but the
background gradually builds up with use.  This increase in background
is mainly due to the 5.3-MeV $\alpha$ from 138-d \nuc{210}{Po} which
comes from the gradual accumulation of 22-year \nuc{210}{Pb}.  With
each use of a cell its \nuc{210}{Pb} content builds up and the
following \nuc{210}{Po} decays produce an ever increasing cell
background.  Each $10^4$~Rn decays increase the cell background rate
by roughly one count per day.  After a cell has been used extensively,
it will eventually have such a high background rate that it must be
retired and replaced with a new one.

The background of each Lucas cell is periodically measured by
evacuating it and counting in the standard manner.  The background of
the cells currently in use in \DO assays is about 10~counts/d and in
the range of (10--20)~counts/d in \HO assays.  These are known with
a 10\% statistical uncertainty.

\subsection{Electronics}

Spurious signals in the photomultiplier dynode chain can be a source
of noise unrelated to alpha decay within a Lucas cell.  The region of
interest for $\alpha$~decays is set to exclude such events, which
mostly occur at low energy.  Nevertheless, there is some leakage into
the region of interest.  By counting without a Lucas cell on the
photomultiplier, this noise rate was measured to be less than 0.5
counts/d.  The lower limit of the region of interest will cut out
some true $\alpha$ events.  Measurements with a Lucas cell spiked with
\nuc{226}{Ra} (which we call a ``hot'' cell) show that the fraction of
events lost depends on the electronics in the counting station and is
in the range of (0--5)\%.  This loss is accounted for in the counting
efficiency uncertainty.

Cross-talk from adjacent counting stations is also a potential source
of noise.  This effect was measured by putting a ``hot'' cell on one
station and measuring the number of counts that appeared in all other
stations.  Even with $\sim 10^7$ counts on the ``hot'' station, the
average number of counts on other stations did not exceed 0.5 counts.
Cross-talk is thus a negligible contributor to the systematic
uncertainty.

The long term photomultiplier gain drift was measured by comparing
``hot'' cell counting rates over a 3-year period.  During this time
the rate in the station used most often for D$_2$O assays varied by
$3.1 \pm 0.9\%$ and the rate in the stations used for H$_2$O assays
varied by no more than $3.5 \pm 2.0\%$.

\section{Data processing}
\label{sec:dataproc}

In this section we derive the relationship between the cumulative
number of detected counts and the concentration $C$ of radon in the
water that enters the degasser.  We assume that $C$ is constant.

If we flow water at a constant rate of $F$ liters/min for a time
interval of duration $t_{\text{assay}}$ and extract with a degassing
efficiency $\epsilon_{\text{degas}}$, then the number of radon atoms
from the water that are present on the first trap in the collection
system (the primary radon trap) at the end of extraction is
 \begin{equation}
 \label{eqn:atomsfromwater}
 N_{\text{water}} = \epsilon_{\text{degas}}\epsilon_{\text{trap}}
                    C F (1 - e^{-\lambda t_{\text{assay}}})/\lambda,
 \end{equation}
where $\epsilon_{\text{trap}}$ is the radon trapping efficiency and
$\lambda$ is the \nuc{222}{Rn} decay constant.  The term in
parentheses is due to the decay of radon during the extraction time; for
short assay times this term is approximately $\lambda
t_{\text{assay}}$.  To obtain the total number of radon atoms present on
the trap we must add to this the background of the radon assay system.
Defining $R_{\text{back}}$ as the rate of radon production by emanation,
diffusion, and ingress in the degasser and the front section of the
collection system, the number of radon atoms from these backgrounds on
the trap at the end of extraction is
 \begin{equation}
 \label{eqn:atomsfrombackground}
 N_{\text{back}} = \epsilon_{\text{trap}} R_{\text{back}}
                   (1 - e^{-\lambda t_{\text{assay}}})/\lambda.
 \end{equation}

These atoms are transferred to the Lucas cell with efficiency
$\epsilon_{\text{transfer}}$.  As discussed in
Sec.~\ref{sec:rncollect}, the radon background of the section of the
collection system used during this transfer is negligible.  If the
time delay between the end of extraction and the start of counting is
$t_{\text{delay}}$, the number of radon atoms present in the Lucas cell
at the start of counting (SOC) is
 \begin{equation}
 \label{eqn:atomsincell}
 N_{\text{SOC}} = \epsilon_{\text{transfer}} e^{-\lambda t_{\text{delay}}}
                  (N_{\text{water}} + N_{\text{back}}).
 \end{equation}

The counting data consist of the superposition of two components:

\begin{itemize}

\item the decay of \nuc{222}{Rn}.  Since the initial number of atoms
is $N_{\text{SOC}}$ and the number exponentially decreases in time,
the count rate of this component varies with time $t$ according to
$\epsilon_{\text{count}}\lambda N_{\text{SOC}} \exp(-\lambda t)$,
where $\epsilon_{\text{count}}$ is the effective \nuc{222}{Rn}
counting efficiency.

\item a constant background rate $B_{\text{Lucas}}$.  These events are
mainly the decay of 22-year \nuc{210}{Pb} which has accumulated in the
cell from previous assays.  The value of $B_{\text{Lucas}}$ is assumed
known from previous counting of this cell for background.

\end{itemize}
We add these two components to get the total count rate as a function
of time
 \begin{equation}
 \label{eqn:countrate}
 \epsilon_{\text{count}} \lambda N_{\text{SOC}} e^{-\lambda t}
   + B_{\text{Lucas}}
 \end{equation}
and integrate for a time interval of duration $t_{\text{count}}$.  The
total number of observed counts in this time is then
 \begin{equation}
 \label{eqn:totalcounts}
 N_{\text{count}} = \epsilon_{\text{count}} N_{\text{SOC}} 
                    (1 - e^{-\lambda t_{\text{count}}})
                    + B_{\text{Lucas}}t_{\text{count}}.
 \end{equation}

Combining Eq. (\ref{eqn:atomsfromwater}),
(\ref{eqn:atomsfrombackground}), (\ref{eqn:atomsincell}), and
(\ref{eqn:totalcounts}), the concentration of radon in the water is
given by
 \begin{eqnarray}
 \label{eqn:conc}
 C = && \frac{ 1 } { \epsilon_{\text{degas}} F } \\ \nonumber
 &&   \times
      [
       \frac{ (N_{\text{count}} - B_{\text{Lucas}} t_{\text{count}}) \lambda }
            { \epsilon (1 - e^{-\lambda t_{\text{assay}}})
              e^{-\lambda t_{\text{delay}}}
              (1 - e^{-\lambda t_{\text{count}}}) }
       -R_{\text{back}}
      ],
 \end{eqnarray}
where we have abbreviated
$\epsilon = \epsilon_{\text{trap}} \epsilon_{\text{transfer}}
            \epsilon_{\text{count}}$.

Typical values of the parameters in this equation are
$\epsilon_{\text{degas}} = 0.6$, $F = 20$~liters/min,
$\epsilon_{\text{trap}} = 1$, $\epsilon_{\text{transfer}} = 0.62$,
$\epsilon_{\text{count}} = 3 \times 0.74$ (there are 3 prompt $\alpha$
particles emitted after each radon decay, each of which is counted
with $\sim$74\% efficiency), $t_{\text{assay}} = 5$~hours in the \DO
system and 30~min in the \HO system, $t_{\text{delay}} = 2$~hours,
$N_{\text{count}} = 400$ from \DO and 740 from \HO,
$t_{\text{count}} = 8$~d, $B_{\text{Lucas}} = 20$~counts/d in the
\HO system and 10 counts/d in the \DO system, and $R_{\text{back}} =
460$~Rn atoms/d in the \HO system and 430~Rn atoms/d in the \DO
system.  To convert $C$ from radon atoms/liter to g~U/g~water,
multiply the right side of Eq.~(\ref{eqn:conc}) by the factor $1.69
\times 10^{-13}/D$ where $D$ is the water density in g/cm$^3$, 1.0 for
\HO and 1.1 for \DO.  This assumes equilibrium in the U decay chain.

\begin{figure}
\begin{center}
\includegraphics*[width=\hsize]{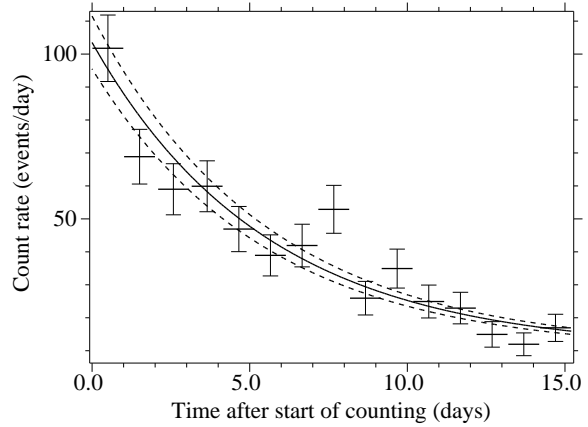} \caption{Count rate
spectrum for radon extracted from \DO.  The background rate for this
cell was $10.0 \pm 0.5$ counts/d.  Data points are indicated by
horizontal lines of 1-d duration, with approximate error limits from
counting statistics.  Since the count rate is very low, the data from
every eight 3-hour data collection intervals have been combined.  The
thick solid line is the count rate predicted from {\protect
Eq.~(\ref{eqn:countrate})} with the encompassing dashed band
indicating the 68\% confidence range from counting statistics.  The
15-d data acquisition time for this spectrum is longer than
customary.}
\label{fig:dataspectrum}
\end{center}
\end{figure}

To check that there is reasonable agreement between the results of
this procedure and the data, the total count rate predicted from
Eq.~(\ref{eqn:countrate}) is calculated and visually compared to the
differential spectrum of number of counts versus counting time.  This
comparison is shown in Fig.~\ref{fig:dataspectrum} for an extraction
from \DO whose activity is slightly higher than average.

\section{Overall systematic uncertainties in the assays}

Most of the various terms that enter the systematic uncertainty have
been given in the previous text and are summarized in
Table~\ref{tab:sys}.  This table also includes the contribution of the
collector, degasser, and cell backgrounds, which will now be
considered.

\begin{table}[ht]
\begin{center}
\caption{Contributions to systematic uncertainty in radon assays in
percent.  All uncertainties are symmetric unless otherwise indicated.
The values for background are evaluated for a water sample whose radon
content is 25\% of the maximum allowable level.}
\label{tab:sys}
\begin{tabular*}{\hsize}{@{} l @{\extracolsep{\fill}} d d @{}}
\hline
\hline
Source                             & \mco{Inner H$_2$O} & \mco{D$_2$O}  \\
\hline
Flow rate $F$                                                   & 14   & 17  \\
Flow rate $F$ (stroke counted)                                  & 10   & 11  \\
Degassing efficiency $\epsilon_{\text{degas}}$                  & 17   & ^{+18}_{-14} \\
Assay system background $R_{\text{back}}$                       &  3   & ^{+10}_{-21} \\
Trapping efficiency $\epsilon_{\text{trap}}$                    &  2   &  2  \\
Transport efficiency to Lucas cell $\epsilon_{\text{transfer}}$ &  3   &  2  \\
Cell counting efficiency $\epsilon_{\text{count}}$              & 10   & 10  \\
Cell background rate $B_{\text{Lucas}}$                         &  7   &  2  \\
Electronic noise                                                & <0.1 & <0.1\\
Photomultiplier stability                                       &  4   &  4  \\
\hline
Combined quadratically             & 26                 & ^{+29}_{-32}  \\
Combined (stroke counted)          & 24                 & ^{+26}_{-30}  \\
\hline
\hline
\end{tabular*}
\end{center}
\end{table}

The systematic uncertainty from backgrounds depends upon the magnitude
of the observed radon signal, which has decreased as \SNO data was
acquired.  On average, during the pure \DO phase of the experiment,
the radon levels in the \HO and \DO were about 1/4 of the maximum
allowable values.  For the inner \HO we thus collected roughly 100
atoms from the water.  The background due to the \HO MDG and the radon
collector for a 30-min assay would result in about $6 \pm 3$~atoms in
the Lucas cell.  Thus, the uncertainty in the system background
contributes a systematic error of $\sim$3\%.  The $^{210}$Pb
background in the Lucas cells is about $20 \pm 2 $~counts/d.  Since in
the first four days of counting we expect $\sim$110~counts from the
water sample, the uncertainty due to cell background is 8/110, or 7\%.


For the \DO, at 1/4 of the maximum allowable level, about 120~Rn atoms
enter the Lucas cell from the water.  The background from the assay
system in a 5-hour assay (the usual assay time) is $56
^{+25}_{-12}$~atoms of radon in the cell.  Thus, the systematic error
due to uncertainty in the assay system background is estimated as
$^{+10}_{-21}\%$.  The $^{210}$Pb background in the Lucas cell
contributes about $8.0 \pm 0.8$~counts/d background.  For the first
four days of counting we expect 135~counts from the radon in the \DO
and thus the systematic error due to uncertainty in cell background is
3/135, or 2\%.


\section{Assay results}

\begin{figure*}
\begin{center}
\includegraphics*[width=0.8\hsize]{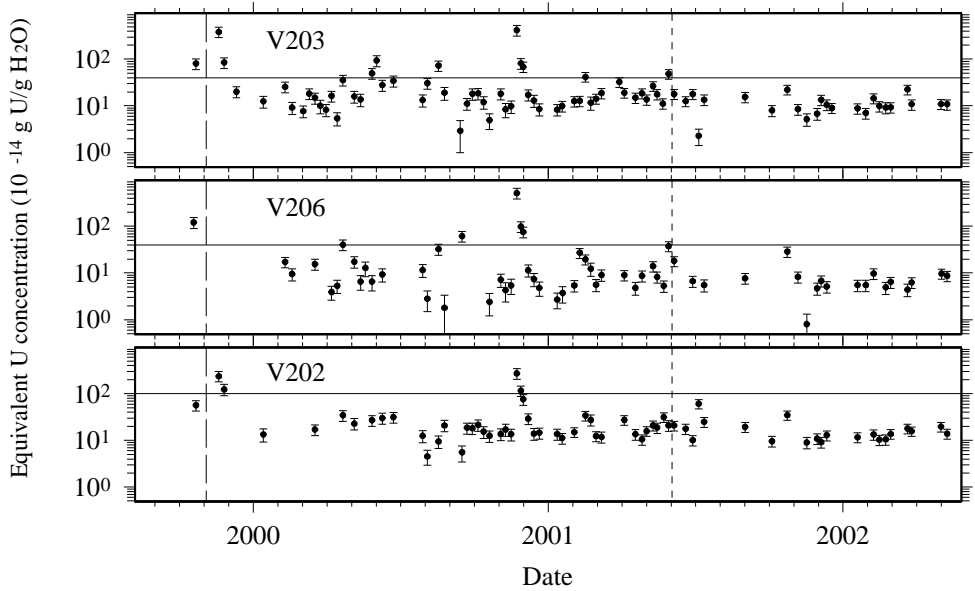}
\caption{Radon levels in H$_2$O.  Three locations are shown:
 V203 at the equator halfway between the acrylic vessel and photomultiplier
 array (upper panel),
 V206 just inside the photomultuplier array at the bottom (middle panel), and
 V202 at the bottom of the cavity containing the bulk shielding
 water (lower panel).
The points show the result of each assay, and the error bars are the
systematic and statistical uncertainty added in quadrature.  The solid
horizontal line denotes the maximum allowable U concentration.  The
dashed line on the left indicates the start of \SNO data acquisition
the dashed line at 2001.4 indicates the time when NaCl was added to
the \DO.  Several unintended radon ``spikes'' of the H$_2$O are seen.
At all three locations the average radon levels are clearly below the
maximum allowable level.}
\label{fig:h2oassay}
\end{center}
\end{figure*}

\begin{figure*}
\begin{center}
\includegraphics[width=0.8\hsize]{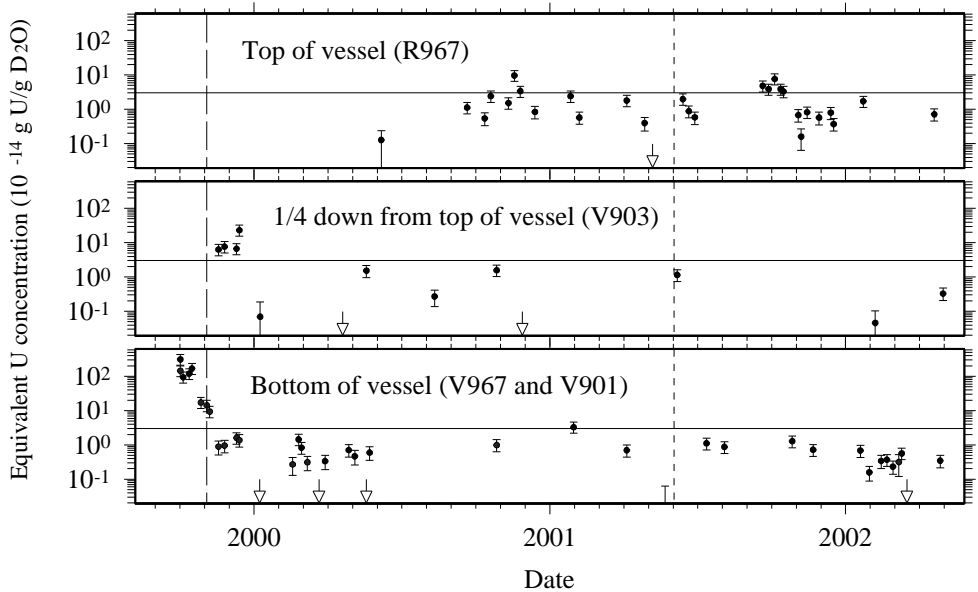} \caption{Radon levels
in D$_2$O.  The points and lines use the same notation as in
Fig.~\protect\ref{fig:h2oassay}.  In some assays, indicated by arrows,
the radon levels measured were negative, due to statistical
fluctuations in the background.  The radon levels in the detector have
averaged about a factor of three below the maximum allowable value.}
\label{fig:d2oassay}
\end{center}
\end{figure*}

Weekly measurements are usually made from the three valves in \HO
shown in Fig.~\ref{fig:sample}, from the \HO process degasser, and
from one or more of the valves in the \DO.  The results of water
assays from 29 September 1999 through 15 May 2002 are shown in
Figures~\ref{fig:h2oassay} and \ref{fig:d2oassay}.  Most of the
scatter in this data is not statistical fluctuation, but is due to
real changes of the radon level in the detector.  For example, there
was a high radon level in both the \HO and \DO during the early period
of neutrino-data taking.  This was because the initially high radon
content of the water used to fill the detector was still decaying.
After that time there are several intervals during which spikes of
radon were unintentionally introduced.  These are clearly apparent in
the H$_2$O plots and were caused by the failure of the H$_2$O process
degasser's vacuum pump, which resulted in laboratory air being
injected into the H$_2$O.  An improved interlock system has prevented
similar faults during subsequent pump failures.  In the first month of
neutrino data-taking, the radon levels in the D$_2$O were still
elevated until the nitrogen ``cover gas'' protection system was
improved to compensate for leaks in the seal of the D$_2$O vapor
space.  After that time, there were two unintentional introductions of
radon into the D$_2$O, but their amplitude was not sufficient to
appreciably interrupt neutrino data-taking.

As is evident by examination of the assay plots, the average radon
levels in both the H$_2$O and D$_2$O are well below the maximum
allowable level.  The interpretation of these assay results in terms
of the net U-chain contamination and the way in which this
contamination influences the neutrino detection rates reported in
\cite{bib:SNOP1,bib:ccnc,bib:dn} will be presented elsewhere.

\section{Summary and Conclusions}

The method that \SNO has developed to determine the radon content of
the water in the detector by direct assay has been described.  This
method is relatively simple to execute, is reliable, and has been
shown to be capable of detecting a few tens of atoms of radon per tonne
of water, equivalent, assuming equilibrium in the U-chain, to a
concentration of a few times $10^{-15}$~g~U/g~water.  This sensitivity
is adequate to show that the \nuc{214}{Bi} contribution to the neutral
current background of the \SNO detector is substantially below the
maximum allowable level, and thus it has been possible for \SNO to
determine the total flux of \nuc{8}{B} solar neutrinos by measurement
of the neutral current interaction rate.

Should the need arise, the sensitivity of this technique could be
improved by an order of magnitude by reducing the background of the
\DO water trap, by constructing new Lucas cells, and by doubling the
assay time.

\section*{Acknowledgements} 

We thank R.~Rodriguez-Jimenez for his aid in calibration and assays.
We thank G.~Carnes, K.~Dinelle, S.~Fostner, B.~McPhail, T.~Spreitzer,
and L.~Wrightson for their contributions to this work during
co-operative student placements with the \SNO experiment.  We
appreciate helpful discussions with A.~Sengupta and J.~Munoz regarding
membrane degasser efficiency.  We are grateful to R.~Lange for
valuable discussions and thank the \SNO operations staff for their
great care in carrying out the assays described here.  We are grateful
to the INCO, Ltd.\ mining company and their staff at the Creighton
mine without whose help this work could not have been conducted and we
greatly thank Atomic Energy of Canada, Ltd.\ (AECL) for the loan of
the heavy water in cooperation with Ontario Power Generation.  This
research was supported in Canada by the Natural Sciences and
Engineering Research Council, the National Research Council, Industry
Canada, the Northern Ontario Heritage Fund Corporation, and the
Province of Ontario; in the USA by the Department of Energy; and in
the United Kingdom by the Science and Engineering Research Council and
the Particle Physics and Astronomy Research Council.  Further support
was provided by INCO, AECL, Agra-Monenco, Canatom, the Canadian
Microelectronics Corporation, AT\&T Microelectronics, Northern
Telecom, and British Nuclear Fuels, Ltd.


\begin{thebibliography}{99}
\bibitem{bib:diaphragm} The assay systems use 1-inch \/ air-operated \/
                        double-diaphragm \/ pumps made by Osmonics Inc.,
                        Minnetonka, Minnesota, USA, that are extensively 
                        modified for vacuum service.
\end{thebibliography}
\end{document}